\documentclass[nofootinbib, reprint, amsmath,amssymb, aps,]{revtex4-1}
\usepackage{graphicx}
\usepackage{dcolumn}
\usepackage{bm}
\usepackage[colorlinks]{hyperref}

\usepackage{graphicx,times}
\usepackage{subfigure}
\usepackage{amsmath}
\usepackage{mathrsfs}
\usepackage{cases}
\usepackage{bm}
\usepackage{longtable}
\usepackage{hyperref}
\usepackage{epstopdf}
\usepackage{amsmath,bm}
\usepackage{amssymb}
\usepackage{natbib}
\usepackage{morefloats}
\usepackage{multirow}
\usepackage{array}
\usepackage{verbatim}
\usepackage[stable]{footmisc}
\usepackage{stackrel}
\usepackage{graphicx}
\usepackage{dcolumn}
\usepackage{bm}

\newcommand{\bx}{{\bf x}}

\begin{document}

\title{Power and spatial complexity in stochastic reconnection}

\author{Amir Jafari}
 \email{elenceq@jhu.edu} 
 
 \author{Ethan Vishniac}
 \email{evishni1@jhu.edu}
 \affiliation {Department of Physics \& Astronomy, Johns Hopkins University, Baltimore, MD, USA}

\author{Vignesh Vaikundaraman}
 \email{v.vaikundaraman@campus.lmu.de}
\affiliation{Ludwig-Maximilians-Universitat, Munich, Germany}

\date{\today}

\begin{abstract}
The level of spatial complexity associated with a given vector field on an arbitrary range of scales can be quantified by a simple, scale-dependent function of time; $0\leq S(t)\leq 1$. Previous work has invoked kinetic and magnetic complexities, associated with velocity and magnetic fields ${\bf u(x}, t)$ and ${\bf B(x}, t)$, to study magnetic reconnection and diffusion in turbulent and magnetized fluids. In this paper, using the coarse-grained momentum equation, we argue that the fluid jets associated with magnetic reconnection events at an arbitrary scale $l$ in the turbulence inertial range are predominantly driven by the Lorentz force ${\bf{N}}_l=({\bf j\times B})_l-{\bf j}_l\times {\bf B}_l$. This force is induced by the subscale currents and is analogous to the turbulent electromotive force ${\cal E}_l=({\bf u\times B})_l-{\bf u}_l\times {\bf B}_l$ in dynamo theories. Typically, high (low) magnetic complexities during reconnection imply large (small) spatial gradients for the magnetic field, i.e., strong (weak) Lorentz forces ${\bf N}_l$. Reconnection launches jets of fluid, hence the rate of change of kinetic complexity is expected to strongly correlate with the power injected by the Lorentz force ${\bf N}_l$. We test this prediction using an incompressible, homogeneous magnetohydrodynamic (MHD) simulation and associate it with previous results. It follows that the stronger (weaker) the turbulence, the more (less) complex the magnetic field and the stronger (weaker) the driving Lorentz forces and thus the ensuing reconnection. 
\end{abstract}

\maketitle

\section{Introduction}\label{Introduction}

Highly electrically conducting fluids threaded by magnetic fields, such as the Earth's outer core or magnetized plasmas in tokamaks and stars, show a remarkable behavior with no counterpart in hydrodynamics: spontaneous, eruptive fluid motions driven by sudden changes in the magnetic field configuration. This phenomenon, called magnetic reconnection, is in fact a ubiquitous feature of turbulent astrophysical fluids. It may also occur in laminar flows, but even in that case reconnection itself can lead to the spontaneous development of turbulence. In any case, it is safe to say that turbulent reconnection is the most prevalent type of reconnection in astrophysical fluids (for a short review of conventional and turbulent reconnection models see \cite{JafariandVishniac2018} and references therein. For a more detailed review of turbulent reconnection see \cite{Lazarian_2019, Review2020}). 

Intense magnetic shears in a non-turbulent fluid correspond to strong electric currents $\bf j=\nabla\times B$, which make the field unstable. The magnetic field can consequently undergo a spontaneous change in its configuration, which may be observed as eruptive fluid motions as the field relaxes to a lower energy state. This Sweet-Parker scheme of reconnection \cite{Parker1957, Sweet1958} can be understood in terms of normal (resistive) diffusion of magnetic field in which the rms separation of magnetic field lines grows linearly with time. In the presence of turbulence, the velocity and magnetic fields will have ill-defined gradients (H{\"o}lder singularity) while the particle trajectories and magnetic field will become stochastic and the normal resistive diffusion of magnetic field will become super-linear Richardson diffusion. Lazarian and Vishniac \cite{LazarianandVishniac1999} developed a successful model of turbulent reconnection based on the idea that random turbulent motions bring initially distant field lines into close separations where they undergo reconnection and super-linearly diffuse away afterwards. This stochastic reconnection model is fast, independent of magnetic diffusivity and in agreement with simulations (\cite{Kowal2009, Kowal2012, Kowal2017}). Later developments illustrated and reformulated it in a mathematically more elaborate fashion, which further showed its intimate connection to Richardson diffusion, field-fluid slippage and stochastic (random) behavior of turbulent magnetic fields \cite{Eyink2011, Eyinketal2013, Eyink2015}).

These developments have been well-established in the last two decades and have been tested numerically (for a detailed review see e.g., \cite{JafariandVishniac2018, Review2020}), yet there remain some vague notions to be clarified mathematically. For instance, it is often said that reconnection can enhance turbulence, or that reconnection helps the reconnecting field relax to a smoother configuration (e.g., see \cite{LazarianandVishniac1999, Lazarianetal2015, JafariandVishniac2018}). These statements may sound physically plausible and mathematically precise. However, a quantitative approach is still in demand in order to formulate notions such as spatially complex or smooth magnetic fields (in a geometric sense) and the enhancement of turbulence by reconnection. The other problem is concerned with the scale-separation in turbulence, i.e., separating the scale of interest into small (turbulent eddy size) and large (usually the system size), which is useful and at the same time also vague, for the definition of small and large scales remain somehow arbitrary. To address these issues, Jafari and Vishniac \cite{JV2019} established a statistical methodology based on a quantitative definition of spatial complexity (or stochasticity) associated with vector fields such as turbulent magnetic and velocity fields. Instead of scale separation, this formalism employes the mathematical concept of distributions (generalized functions) to coarse-grain magnetohydrodynamic (MHD) equations in the turbulence inertial range. Later, this formalism was tested numerically and used to approach stochastic magnetic reconnection \cite{SecondJVV2019} and also magnetic diffusion in turbulence \cite{JVV2019}. In the present paper, based on these recent developments, briefly reviewed in \S\ref{complexities}, we investigate the turbulent, Lorentz forces which drive stochastic magnetic reconnection. 

As for the detailed plan of this paper, we start in \S \ref{complexities}, with a brief review of the previous developments (\cite{JV2019, SecondJVV2019}) regarding kinetic and magnetic complexities and their role in the study of turbulent magnetic fields and reconnection. In \S\ref{reconnection}, we present arguments based on the coarse-grained induction equation to quantify already known result that reconnection is driven by different phenomena at different scales. In \S \ref{recfield}, using the coarse-grained form of the Navier-Stokes equation, we derive the exact form of the Lorentz force driving stochastic reconnection and clarify its role in local slippage and reconnection events in MHD turbulence. We define the reconnection power $\cal P$, in terms of the Lorentz force $\bf{N}$, and also test our results using an incompressible, homogenous numerical simulation from Johns Hopkins Turbulence Database archived online (\cite{JHTDB, JHTB1, JHTB2}). This is a direct numerical simulation (DNS), using $1024^3$ nodes, which solves incompressible MHD equations using pseudo-spectral method. The simulation time is $2.56$ (code units) and $1024$ time-steps are available (the frames are stored at every $10$ time-steps of the DNS). Energy is injected using a Taylor-Green flow stirring force. We check our results in randomly selected sub-volumes of the simulation box with different sizes. Dividing the simulation box into sub-volumes of different sizes, and performing the computations separately in each sub-volume, reduces the computation time and also allows us to test the theoretical predictions in different regions of the box as statistical samples. In \S\ref{summary}, we summarize and discuss our results.


\section{Kinetic and Magnetic Complexities}\label{complexities}

In this section, we briefly review the implications of magnetic and kinetic spatial complexities for magnetic reconnection; for more details see \cite{JV2019, SecondJVV2019}. We will invoke these concepts in \S\ref{recfield} in order to relate the Lorentz force and reconnection power to the rate of change of the kinetic complexity.

In a fluid with the velocity field ${\bf u}$, a fluid parcel of an arbitrary size $l$ located at spacetime point $({\bx}, t)$ has an average velocity 

\begin{equation}\label{coarsegrain1}
{\bf{u}}_l ({\bf{x}}, t)=\int_V G\Big({{\bf r}\over l}\Big)  {\bf u}({\bf{x+r}}, t) {d^3r\over l^3},
\end{equation}
where $G({\bf r}/l) = G(r/l) $ is a smooth and rapidly decaying kernel, e.g., $G({\bf r}/l)\sim e^{-r^2/l^2}$ . The real, mathematical field $\bf u$ is sometimes called the bare field while ${\bf u}_l$ is called the renormalized, or coarse-grained, field at scale $l$.\footnote{For simplicity, one may also assume $G({\bf{r}})\geq 0$, $\lim_{|\bf r|\rightarrow \infty} G({\bf{r}})\rightarrow 0$, $\int_V d^3r G({\bf{r}})=1$, $\int_V d^3r \; {\bf{r}}\;G({\bf{r}})=0$, $\int_V d^3r |{\bf{r}}|^2 \;G({\bf{r}})= 1$ and $G({\bf{r}})=G(r)$ with $|{\bf{r}}|=r$. } Differential equations containing the velocity field $\bf u$, or magnetic field $\bf B$, can also be multiplied by the kernel $G$ and integrated to get the corresponding renormalized equations. The renormalized form of the momentum equation for example reads

\begin{equation}\label{NS3}
{\partial {\bf{u}}_l\over \partial t}=({\bf j}\times {\bf B})_l-\nabla . ({\bf{u}}{\bf{u}})_l-\nabla p_l+\nu\nabla^2{\bf u}_l,
\end{equation}
where ${\bf j}_l=\nabla\times {\bf B}_l$ is the coarse-grained electric current, $p_l$ is the coarse-grained pressure field, and $\nu$ denotes viscosity.

 As mentioned before, the coarse-grained field ${\bf u}_l({\bx}, t)$ is the weighted-average velocity  field of a fluid parcel of size $l$ at point $\bx$. Since $G(r/l)$ is a rapidly decaying function so the integral ${\bf u}_l({\bf{x}}, t)=l^{-3}\int_V G({\bf r}/l)  {\bf u}({\bf{x+r}}, t) d^3r$ gets smaller and smaller contributions from points at distances $\gg l$ from $\bf x$. If we renormalize the field at a larger scale $L>l$, on the other hand, we will get the average velocity field of a fluid parcel of scale $L$ at the same point $\bx$, i.e., ${\bf u}_L({\bx}, t)$ which will generally differ from ${\bf u}_l({\bx}, t)$. In a laminar flow whose velocity field has a large curvature radius $\gg L$ (e.g., a very simple, smooth pipe flow), we expect $\hat{\bf u}_l.\hat{\bf u}_L\simeq 1$. For a stochastic velocity field in a fully turbulent medium, on the other hand, $-1\leq \hat{\bf u}_l.\hat{\bf u}_L\leq 1$ becomes a rapidly varying stochastic variable. This quantity in fact measures the spatial complexity (or stochasticity level) of $\bf u$ at point $\bf x$. Its root-mean-square (rms) value tells us how spatially complex (or stochastic) the flow (i.e., the velocity field) is on average in a given volume $V$. In order to obtain a non-negative global quantity, we can volume average ${1\over 2}|\hat{\bf u}_l({\bx}, t).\hat{\bf u}_L({\bx}, t)-1|$. Using root-mean-square (rms) averaging, for instance, the spatial complexity, or stochasticity level, of the velocity field $\bf u$ in an arbitrary spatial volume $V$ can be defined as (\cite{JV2019, JVV2019, SecondJVV2019}):

\begin{eqnarray}\label{magnetic}
S_k(t)&=&{1\over 2} (\hat{\bf u}_l.\hat{\bf u}_L-1)_{rms}\\\notag
&=&{1\over 2} \Big(  \int_V|\hat{\bf u}_l.\hat{\bf u}_L-1|^2  {d^3x\over V} \Big)^{1/2},
\end{eqnarray}
Magnetic complexity, or stochasticity level for a turbulent field, is defined similarly;

\begin{equation}\label{kinetics}
S_m(t)={1\over 2}( \hat{\bf B}_l.\hat{\bf B}_L-1)_{rms},
\end{equation}

 \begin{figure}
 \begin{centering}
\includegraphics[scale=.55]{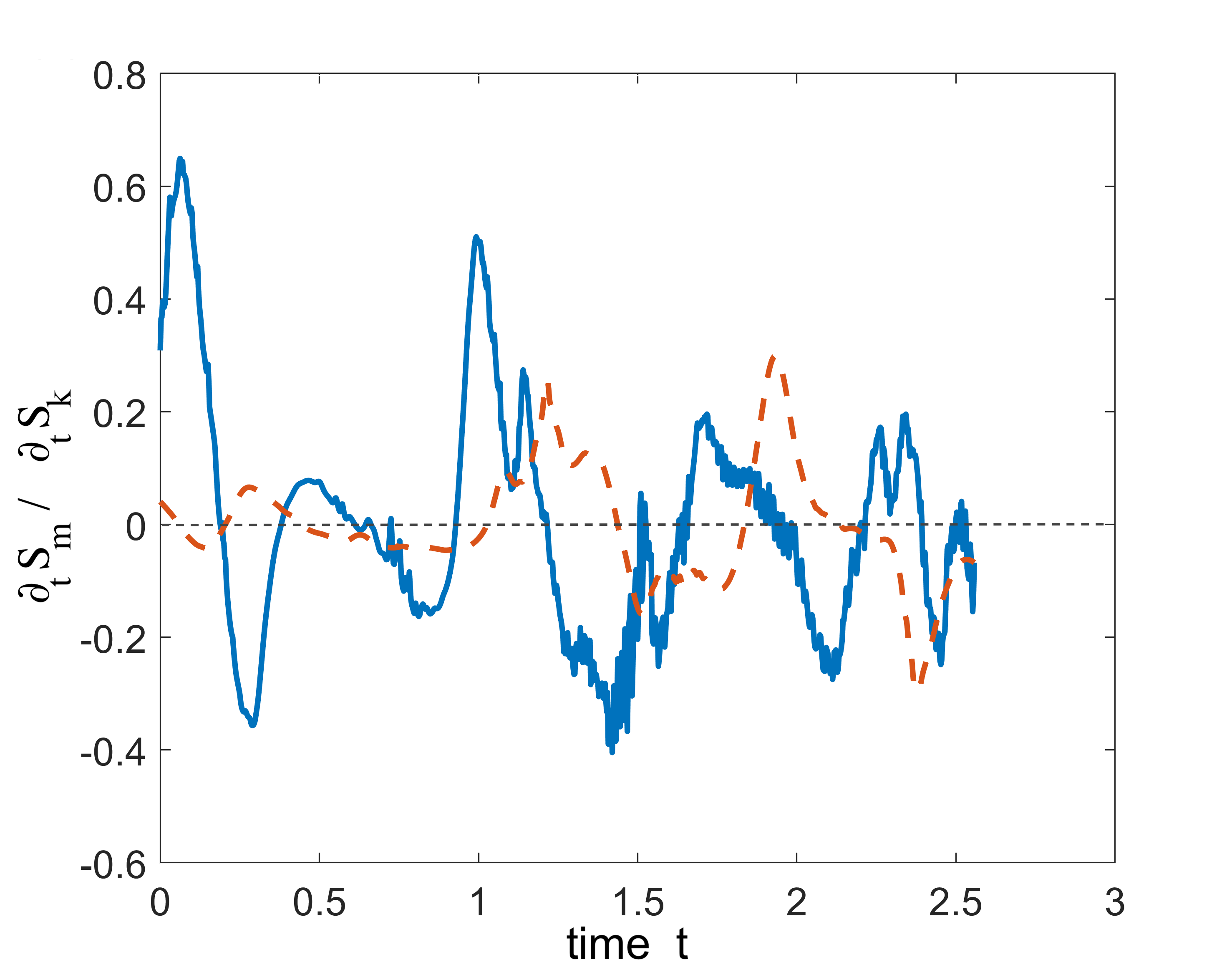}
\caption {\footnotesize {Rate of change of magnetic complexity $\partial_t S_m$ (solid, blue curve) and kinetic complexity $\partial_t S_k$ (dashed, red curve) in a sub-volume of the simulation box where magnetic energy dissipates efficiently. Local maxima of magnetic complexity (points where the solid, blue curve vanishes with a negative slope; $\partial_t S_m=0$ \& $\partial_t^2 S_m<0$) implies the onset of field's slippage through the fluid. If magnetic energy dissipates efficiently, the relaxing field may also accelerate fluid elements increasing the kinetic complexity. This corresponds to local magnetic reversals, observed in this graph. The positive values of $\partial_t S_k$ right after these local maxima of $S_m$ implies increasing kinetic complexity $\partial_t S_k>0$, corresponding to eruptive fluid motions. Hence, the presence of this correlation between $\partial_t S_m$ and $\partial_t S_k$ in any region is interpreted as a sign of magnetic reversals \cite{SecondJVV2019}. }}\label{zap2}
\end{centering}
\end{figure}

 This simple statistical formalism proves very useful in problems such as magnetic reconnection. In a turbulent, magnetized fluid, the magnetic field fails to follow particle trajectories (i.e., failure of Alfv\'en flux freezing) but it can still follow the flow in a statistical sense (statistical flux freezing; see \cite{Eyink2015}). Thus the field will still become tangled by the turbulent flow, thus its spatial complexity $S_m$ will increase correspondingly until the magnetic tension forces make the field slip through the fluid, lowing its complexity. This corresponds to a local maximum for the function $S_m(t)$ corresponding to conditions $\partial_t S_m=0$ \& $\partial_t^2 S_m<0$. If the slipping field can effectively accelerate fluid particles, the magnetic energy will be efficiently converted into kinetic energy. The resultant eruptive fluid motions increase the spatial complexity of the velocity field $S_k(t)$. This is magnetic reconnection, characterized by positive values of $\partial_t S_k$ (eruptive motions) right after $S_m$ reaches its local maxima. As magnetic field relaxes, i.e, $S_m$ decreases, the eruptive motions increase $S_k$ and this in turn enhances turbulence which tends to tangle the field once again and increase $S_m$. Therefore, we expect $\partial_t S_k$ to be followed by $\partial_t S_m$; see Fig.(\ref{zap2}). This picture seems to be in agreement with magnetohydrodynamic simulations (\cite{JV2019, SecondJV2019}). In passing, let us also emphasize that in fact the statistical formalism briefly discussed above is based on a single scalar field called scale split energy density, which for magnetic and velocity fields is defined as
 
\begin{equation}\label{scale-split-energy}
\begin{cases}
\psi_{lL}({\bx}, t)={1\over 2}{\bf B}_l.{\bf B}_L,\\
\Psi_{lL}({\bx}, t)={1\over 2}{\bf u}_l.{\bf u}_L,\\
\end{cases}
\end{equation}

which can be written in terms of two scalar fields as $\psi_{lL}=\chi_{lL}\phi_{lL}$ (or $\Psi_{lL}={\cal X}_{lL}\Phi_{lL}$ for the velocity field $\bf u$). Here $\phi_{lL}=\hat{\bf B}_l.\hat{\bf B}_L$ ($\Phi_{lL}=\hat{\bf u}_l.\hat{\bf u}_L$) is associated with magnetic (kinetic) topology used to define spatial complexity (or stochasticity level) $S_m(t)={1\over 2}(\phi_{lL}-1)_{rms}$ ($S_k(t)={1\over 2} (\Phi_{lL}-1)_{rms}$) whereas $\chi_{lL}={1\over 2}B_l B_L$ (${\cal X}_{lL}={1\over 2}u_l u_L$) is associated with magnetic (kinetic) energy and is in fact the geometric mean of the energy densities at scales $l$ and $L$; $\chi_{lL}=\sqrt{ {B_l^2\over 2} .{B_L^2\over 2}}$ (${\cal X}_{lL}=\sqrt{ {u_l^2\over 2} .{u_L^2\over 2}}$); see \cite{JV2019}.

In short, stochastic flux freezing \cite{Eyink2011} implies that magnetic field will follow a turbulent flow in a statistical sense, therefore, we expect that complicated, random motions in turbulence will increasingly entangle the magnetic field in a complex way \cite{JV2019}. This qualitative statement can be translated into a more quantitative statement by introducing the magnetic and kinetic complexity functions, $S_m$ and $S_k$ \cite{SecondJVV2019}. How does a geometrically complex magnetic field, associated with a large $S_m$, in turn increase the kinetic complexity $S_k$ of the flow while relaxing to a smoother magnetic configuration? The answer to this question can be quantified by means of the Lorentz forces.

\section{Reconnection at Different Scales}\label{reconnection}

In this section, following \cite{JV2019}, we show how the evolution of the unit, direction vector $\hat{\bf B}\equiv {\bf B}/B$ at an arbitrary scale $l$ can be used to obtain useful information about reconnection and field-fluid slippage on that scale.

During a reconnection event, by definition, the direction of the magnetic field undergoes rapid changes. Hence, let us focus on the direction, rather the magnitude, of the coarse-grained magnetic field  ${\bf B}_l$ at the inertial scale $l$ using the unit, magnetic direction vector $\hat{\bf B}_l={\bf B}_l/B_l$, with $B_l=|{\bf B}_l|$, at scale $l$. It is governed by 
\begin{eqnarray}\label{Bdirection}
  {\partial \hat {\bf{B}}_l\over \partial t}&-&{\nabla\times({\bf u}_l\times{\bf B}_l)_\perp \over B_l}=\underbrace{{(\nabla\times{\cal E}_l)_\perp\over B_l}}_\text{turbulent EMF}\\\notag
  &-&\underbrace{   \eta{(\nabla\times{\bf j}_l)_\perp\over B_l}}_\text{resistivity}-\underbrace{{1\over B_l }\nabla\times\Big({{\bf{N}}_l\over ne}+{{\bf B}_l.\nabla{\bf B}_l\over ne }\Big)_\perp}_\text{Hall effect}+\dots\;,\\\notag
  \end{eqnarray}
where we have only kept the turbulent non-linear term, the resistive electric field and Hall effect (it is simple calculus to derive eq.(\ref{Bdirection}); see \cite{JV2019}). Here, $(\;)_\perp$ denotes the perpendicular component with respect to ${\bf B}_l$; $n$ and $e$ denote respectively the electron density and charge; $\eta$ is magnetic diffusivity and
\begin{equation}\label{EMF}
{\bf \cal{E}}_l=({\bf u\times B})_l-{\bf u}_l\times{\bf B}_l
\end{equation}

is the turbulent electromotive force (EMF). The vector field 

\begin{equation}\label{Rfield}
{\bf{N}}_l=({\bf j\times B})_l-{\bf j}_l\times{\bf B}_l
\end{equation}

represents the average internal Lorentz forces acting on a parcel of size $l$, which appears here through the Hall term. Interestingly, the source terms in the RHS of eq.(\ref{Bdirection}) also drive the relative velocity (when appropriately defined) between the filed and fluid, which was shown by Eyink \cite{Eyink2015} to be intimately related to field-fluid slippage and stochastic magnetic reconnection. Note that during a reconnection event, one or more of the source terms present in the RHS of eq.(\ref{Bdirection}) should drive rapid changes in $\hat{\bf B}_l$.

In the inertial range of turbulence, the non-linear turbulent term $(\nabla\times{\cal E}_l)_\perp/B_l$ will dominate all the remaining plasma effects \cite{Eyink2015}. As another example, in a region of scale $l$ in a non-turbulent fluid, where oppositely directed magnetic vectors give rise a strong electric current, the term $\sigma_{l\text{res}}=\eta|(\nabla\times{\bf j}_l)_\perp|/B_l$ can be large such that it dominates over other terms. This corresponds to the Sweet-Parker reconnection. Mathematically, if the RHS of eq.(\ref{Bdirection}) vanishes identically (which is merely a mathematical rather than a physical assumption), the magnetic field will follow the velocity field (standard Alfv\'en flux freezing in laminar flows). However, the non-linear turbulent term $(\nabla\times{\cal E}_l)_\perp/B_l$ as well as the remaining non-ideal plasma terms in eq.(\ref{Bdirection}) act as source terms in this differential equation driving the field-fluid slippage---spontaneous changes in magnetic field direction $\hat{\bf B}_l$ (\cite{Eyink2015}; \cite{JV2019}). With $l$ in the turbulence inertial range, the plasma terms in eq.(\ref{Bdirection}) can be neglected \cite{JV2019}, hence the field-fluid slippage will be driven entirely by the turbulence.

\section{Lorentz Force}\label{recfield}

In this section, we present the main idea of this paper, that the Lorentz force ${\bf N}_l$ is the force field driving reconnection at an inertial scale $l$ and its corresponding power (rate of energy injection by the field) is correlated with the rate of change of kinetic complexity (due to the magnetically driven fluid jets). 

In order to obtain an expression for the Lorentz forces driving stochastic reconnection, let us write the momentum equation;

\begin{eqnarray}\notag
{\partial_t {\bf{u}}}\;+\;\nabla . ({\bf{u}}{\bf{u}})\;=\;{\bf j}\times {\bf B}\;-\nabla p\;+\;\nu\nabla^2{\bf u},
\end{eqnarray}
 and compare it with its coarse-grained version;
\begin{eqnarray}\label{NS31}
{\partial_t {\bf{u}}_l}+\nabla . ({\bf{u}}_l{\bf{u}}_l)&=&{\bf j}_l\times {\bf B}_l-\nabla p_l+\nu\nabla^2{\bf u}_l\\\notag
&+&\underbrace{{\bf{N}}_l-\nabla.{\bf{\cal M}}_l}_\text{non-linear turbulent terms},
\end{eqnarray}
where 

$${\bf{\cal M}}_l=({\bf uu})_l-{\bf u}_l{\bf u}_l$$

is the turbulent stress tensor. These expressions show how coarse-graining introduces two non-linear terms ${\bf{N}}_l$ and ${\bf{\cal M}}_l$, which generally are not negligible in the presence of turbulence. The term ${\bf j}_l\times{\bf B}_l$, as the counterpart of $\bf j\times B$, is the Lorentz force acting on the parcel of fluid on scale $l$. However, there is a residue; the internal Lorentz forces, arising from the turbulent motions inside the parcel, add up to ${\bf{N}}_l$. Unlike the Lorentz force ${\bf j}_l\times{\bf B}_l$, which tries to move the whole parcel, the turbulent force ${\bf{N}}_l$ is a kind of internal force. The analogy between the electromotive force ${\bf \cal{E}}_l$, eq.(\ref{EMF}), which is the electric field induced by the turbulent motions below the scale $l$, and the Lorentz force ${\bf{N}}_l$, eq.(\ref{Rfield}), suggests that ${\bf{N}}_l$ is generated by turbulent currents below the scale $l$. The resultant small scale, magnetically driven fluid motions are related to local reconnection events predicted in the stochastic reconnection model (\cite{LazarianandVishniac1999}; \cite{JafariandVishniac2018}).

Physically, a large magnetic complexity translates into strong magnetic shears and thus intense electric currents $\bf j=\nabla\times B$. This in turn can generate appreciable magnetic forces $\bf j\times B$, which can potentially enhance turbulence. Generally, the more spatially complex the magnetic field is, the stronger the Lorentz forces are expected to be. This in turn means an efficient acceleration of fluid jets. The contribution of the foce ${\bf {N}}_{l}$ to the changes in the kinetic spatial complexity, by means of injecting energy to the flow, is given by \cite{SecondJV2019} 

\begin{eqnarray}\label{Tdeform2}
{\partial S_k(t)\over\partial t}\Big|_{rec}={1\over 2} \int_V&&{d^3x\over V}{\hat{\bf u}_l.\hat{\bf u}_L-1  \over (\hat{\bf u}_l.\hat{\bf u}_L-1)_{rms}} \\\notag
&&\times {1\over u_lu_L} \Big ({\bf u}_L\;.\;  {{\bf{N}}_l}_\perp +{\bf u}_l\;.\; {{\bf{N}}_L}_\perp  \Big),
\end{eqnarray}

where ${{\bf{N}}_l}_\perp$ (${{\bf{N}}_L}_\perp$) is the perpendicular component of ${\bf{N}}_l$ (${\bf{N}}_L$) with respect to ${\bf u}_l$ (${\bf u}_L$). The terms inside the parentheses in eq.(\ref{Tdeform2}) represent the power. To see this more clearly, we can use eq.(\ref{NS31}) to write

\begin{eqnarray}\notag
{\partial\Psi_{lL}\over\partial t}&=&{1\over 2} \Big({\bf u}_l\;.\; {\bf{N}}_L     \;+\;{\bf u}_L\;.\; {\bf{N}}_l\Big)\\\notag
&&+{1\over 2}\Big[{\bf u}_l\;.\;({\bf j}_L\times{\bf B}_L)+{\bf u}_L\;.\;({\bf j}_l\times{\bf B}_l)\Big]\\\label{dtpsi}
&&+\;\text{(non-magnetic terms)},\\\notag
\end{eqnarray}

where $\Psi_{lL}={1\over 2}{\bf u}_l.{\bf u}_L$ is the kinetic scale-split energy defined by eq.(\ref{scale-split-energy}). 
Here, the term ${\bf j}_l\times{\bf B}_l$ (or ${\bf j}_L\times{\bf B}_L$) represents the Lorentz force exerted on the whole parcel of size $l$ ($L$) which is generated by the electric current ${\bf j}_l=\nabla\times {\bf B}_l$ (${\bf j}_L=\nabla\times {\bf B}_L$) at the scale $l$ ($L$). On the other hand, the term ${\bf{N}}_l$ (${\bf{N}}_L$) is the average magnetic force generated by the currents below the scale $l$ ($L$), corresponding to the internal magnetic reversals inside the parcel. Therefore, for the rate at which these local reversals inject energy into the flow, we define

\begin{equation}\label{main1}
{\partial\Psi_{lL}\over\partial t}\Big|_{rec} \equiv{1\over 2}\Big({\bf u}_l\;.\; {\bf{N}}_L+{\bf u}_L\;.\; {\bf{N}}_l\Big).
\end{equation}

The RHS of the above expression is a measure of energy injection rate (power) at a range of scales, hence by averaging we find

$$
{\cal P}_{lL}= \Big(\Big[{\partial\Psi_{lL}\over\partial t} \Big]_{rec} \Big)_{rms},$$

where we have defined the reconnection power as
\begin{equation}\label{Power}
{\cal P}_{lL}\equiv{1\over 2} \Big({\bf u}_l.{\bf{N}}_L+{\bf u}_L.{\bf{N}}_l\Big)_{rms}.
\end{equation}

Therefore, if magnetic reconnection dominates over other effects in a given region, we can expect 

\begin{eqnarray}\label{secondNS31second}
{\cal P}_{lL}\sim \Big(\Big[{\partial\Psi_{lL}\over\partial t} \Big]\Big)_{rms}.
\end{eqnarray}

\begin{figure}
 \begin{centering}
\includegraphics[scale=.55]{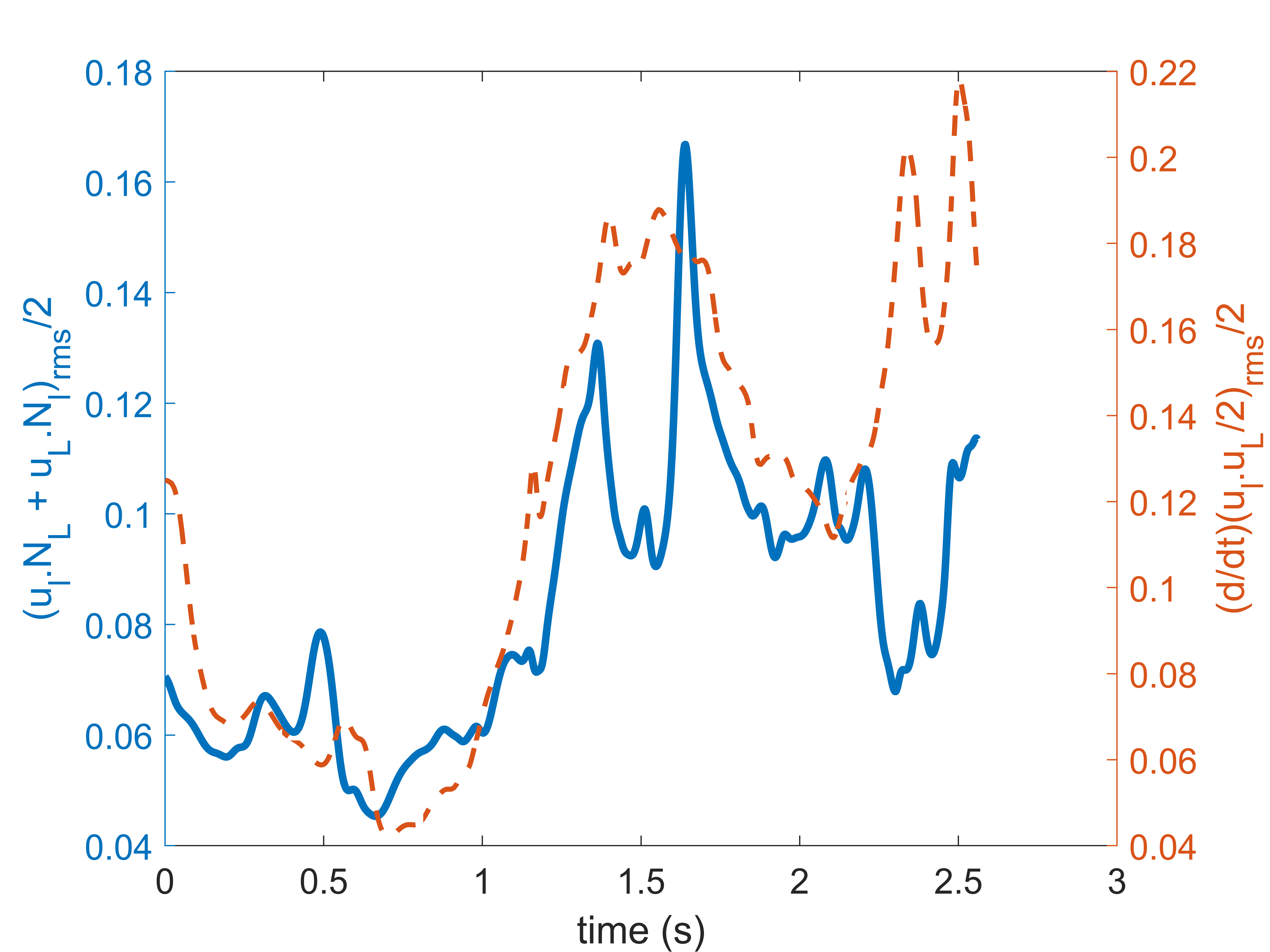}
\caption {\footnotesize {Reconnection power ${\cal P}_{lL}$ (solid, blue curve) and $(\partial_t\Psi_{lL})_{rms}$ (dashed, red curve) in the same region of the simulation box as the one in Fig.(\ref{zap2}). These two functions are strongly correlated (with a typical cross-correlation above $0.90$) in regions where magnetic reversals seem to be strong, i.e., reconnection zones where $(\partial_t\Psi_{lL})_{rms}$ is  dominated by $([\partial_t\Psi_{lL}]_{rec})_{rms}\equiv{\cal P}_{lL}$. This means the dominance of the terms inside the first parentheses in the RHS of (\ref{dtpsi}) when averaged, i.e., the relationship (\ref{secondNS31second}) holds within a good approximation. In other regions, where the correlation shown in Fig.(\ref{zap2}) does not seem to hold implying the absence of magnetic reversals, this correlation fades away too, implying the dominance of non-magnetic effects in driving $\partial_t\Psi_{lL}$; see eq.(\ref{dtpsi}).  }}\label{zap10}
\end{centering}
\end{figure}

In Fig.(\ref{zap10}), we have plotted $(\partial_t\Psi_{lL})_{rms}$ and ${\cal P}_{lL}$ (for $l=7, L=21$ in grid units) in the same region of the simulation box as used in Fig.(\ref{zap2}). Comparing the cross-correlations between these two time series shows a strong correlation with lag zero, i.e., the two functions are strongly correlated when compared without shifting any of them in time (with a cross-correlation above $0.92$). In other words, in regions where reversals seem to be present as Fig.(\ref{zap2}) indicates \cite{SecondJVV2019}, the terms inside the first parentheses in the RHS of eq.(\ref{dtpsi}) dominate the other terms. This behavior is typically seen other in reconnection regions while it is insensitive to the choice of the inertial scales $l$ and $L$. This picture suggests that $\bf N$ plays a major role in driving reconnection. In fact, although at some regions of the box, the correlation between the reconnection power $\cal P$ and the rate of change of kinetic complexity $\partial_t\Psi$ is much stronger compared with other regions, however, overall we see a positive correlation between these quantities almost everywhere in MHD turbulence. This is in accordance with the fact that even in the absence of large scale reconnection events, small scale magnetic reversals are an essential part of MHD turbulence \cite{SecondJVV2019}.

To put the previous findings in the perspective, note that $\partial_t S_k(t)$ has already been shown \cite{SecondJVV2019} to be statistically correlated with the rate of change of magnetic complexity $\partial_t S_m(t)$ (see Fig.(\ref{zap2}) in \S\ref{complexities}). A local maximum for $S_m$, defined by the conditions $\partial_t S_m=0$, $\partial^2_t S_m<0$, implies the occurrence of field-fluid slippage. It corresponds to a highly spatially complex magnetic field relaxing to a smoother configuration. If the relaxing field is intense enough in a large volume of fluid that it can accelerate the fluid elements, the kinetic spatial complexity will increase consequently, i.e., $\partial_t S_k>0$ which may correspond to reconnection \cite{JVV2019}. The field ${\bf{N}}_l$ is the force that drives the reconnection and injects energy into the flow ${\bf u}_l$, therefore, the power or the energy conversion per unit time is of order of ${\bf u}_l.{\bf N}_l$. A parcel of size $l$ moves with velocity ${\bf u}_l$ in the force field ${\bf N}_L$ generated by the fluid elements contained in a larger parcel of size $L$. Likewise, the larger parcel of size $L>l$ moves with the velocity ${\bf u}_L$ while it experiences the internal forces ${\bf N}_l$. This provides an intuitive way to interpret the definition of power given by eq.(\ref{Power}).

 \section{Summary and Conclusions}\label{summary}
 
Alfv\'en theorem states that the magnetic fields threading highly electrically conducting fluids tend to freeze into the flow, approximately following the particle trajectories. In turbulence, however, the field follows the turbulent flow only in a statistical sense (referred to as stochastic flux freezing put forward by Eyink \cite{Eyink2011}). Turbulence will tangle an initially smooth magnetic field, gradually increasing its spatial complexity $S_m$. Due to the build-up of magnetic tensions, however, $S_m$ will at some point reach a local maximum and the field will start to slip through the fluid to relax to a smoother configuration with a lower spatial complexity. This field-fluid slippage, corresponding to a maximum $S_m$, may also accelerate the fluid elements converting magnetic energy into kinetic energy---magnetic reconnection---thereby increasing the complexity of the flow, i.e., $S_k$. In fact, highly complex magnetic configurations, e.g., on the solar surface, seem to be followed by highly complex, eruptive fluid motions, by means of which magnetic field relaxes to a smoother configuration. Turbulence starts to tangle this smooth field once again, hence the cycle more or less repeats itself. In order to study this picture statistically, we can use the magnetic and kinetic spatial complexities $S_m(t)$ and $S_k(t)$ (which can be treated as time series in numerical computations). Previous work (\cite{JV2019}; \cite{SecondJVV2019}; \cite{JVV2019}; \cite{dynamicsJV2019}) has investigated different aspects of this theoretical picture using both analytic and numerical methodologies in terms of magnetic topology, kinetic and magnetic spatial complexities and energies. It goes without saying, of course, that MHD turbulence is much more complicated than what such a simplified picture based on a few time series may imply, however, as far as a statistical understanding is desired, they might prove useful in the study of turbulent magnetic fields.

In this paper, we have presented physical arguments, supported by the results of an incompressible, homogeneous MHD simulation, in order to show that the vector field ${\bf{N}}_l=({\bf j\times B})_l-{\bf j}_l\times{\bf B}_l$ plays an important role in turbulent magnetic reconnection in the turbulence inertial range. Reconnection power defined as ${1\over 2}({\bf u}_l.{\bf{N}}_L+{\bf u}_L.{\bf{N}}_l)_{rms}$ is a measure of the rate at which the reconnecting magnetic field injects energy to the flow. As the fluid jets are launched by the highly tangled (i.e., large magnetic complexity), reconnecting magnetic field, the kinetic complexity increases. High magnetic complexities imply large magnetic shears and strong Lorentz forces $\bf N$. Therefore, if we take the rate at which $\bf N$ injects energy to the flow proportional to $\bf u.N$ on a range of inertial scales, it should be correlated with the rate at which kinetic complexity changes, i.e., $\partial_t\Psi_{lL}$. Numerical simulations seem to be in agreement with this picture, although more comprehensive numerical studies are necessary to reach a conclusive picture. This picture is also complimentary to the previous findings according to which the particular relationship between $\partial_t S_m$ and $\partial_t S_k$, discussed in \S\ref{complexities} and seen in Fig.(\ref{zap2}), implies strong magnetic slippage or reconnection events in the corresponding spatial volume. If so, we expect $(\partial_t\Psi_{lL})_{rms}$ to be mostly driven by the magnetic interactions, i.e., ${\cal P}_{lL}$. It turns out that in regions where magnetic reversals are strong enough to efficiently dissipate magnetic energy and drive fluid jets (which give rise to the particular relationship between $\partial_t S_m$ and $\partial_t S_k$ discussed in \S \ref{complexities}), the Lorentz force $\bf N$ seems to drive the reconnection and the relationship given by (\ref{secondNS31second}) is satisfied.

The data that support the findings of this study are openly available in the Johns Hopkins Turbulence Database at http://turbulence.pha.jhu.edu, \cite{JHTDB, JHTB1, JHTB2}.

\bibliographystyle{apsrev4-2}
\bibliography{RecField3}

\end{document}